\documentclass[preprint,showpacs,preprintnumbers,amsmath,amssymb,aps,nofootinbib]{revtex4}
\usepackage{bm}
\usepackage{amsfonts}
\usepackage{amssymb}
\usepackage{graphicx}
\usepackage{color}
\begin{document}

\title{Polarized Deep Inelastic  Scattering Off  the ``Neutron''\\  From Gauge/String Duality}
\author{Jian-Hua Gao}
\email{gaojh79@ustc.edu.cn}
\affiliation{Department of Modern Physics, University of Science and Technology of China,
Hefei, Anhui 230026, People's Republic of China}
\author{Zong-Gang Mou}
\email{mouzonggang@mail.sdu.edu.cn}
\affiliation{Department
of Physics, Shandong University, Jinan, Shandong, 250100, People's
Republic of China}
\date{\today}

\begin{abstract}
We investigate deep inelastic scattering off the polarized
``neutron'' using gauge/string duality. The ``neutron'' corresponds
to a supergravity mode of the neutral dilatino. Through introducing
the Pauli interaction term into the action in $\textrm{AdS}_{5}$
space, we calculate the polarized deep inelastic structure functions
of the ``neutron''  in supergravity approximation at large t' Hooft
coupling $\lambda$ and finite $x$ with $\lambda^{-1/2}\ll x<1$. In
comparison with the charged dilatino ``proton,'' which has been
obtained in the previous work by Gao and Xiao, we find the structure
functions of the ``neutron'' are power suppressed at the same order
as the ones of the ``proton.''  Especially, we find the
Burkhardt-Cottingham-like sum rule, which is satisfied in the work
by Gao and Xiao, is broken  due to  the Pauli interaction term. We
also illustrate how such a Pauli interaction term can arise
naturally from  higher dimensional  fermion-graviton coupling
through the usual  Kaluza-Klein reduction.
\end{abstract}

\pacs{11.25.Tq, 13.88.+e, 13.60.Hb}

 \maketitle
\section{Introduction}
Gauge/string
duality\cite{Maldacena:1997re,Witten:1998qj,Gubser:1998bc} provides
us with new insights into gauge theories in a strong coupling
regime. There have been substantial progresses in studying strong
coupling gauge theories by using such gauge/string duality. A few
years ago, Polchinski and Strassler\cite{Polchinski:2001tt,
Polchinski:2002jw} studied the deep inelastic scattering on hadrons
by using gauge/string duality, in which the spinless hadron and
spin-$\frac{1}{2}$ hadron correspond to supergravity modes of
dilaton and dilatino, respectively. The usual structure functions
$F_1$ and $F_2$ are calculated for both spinless and
spin-$\frac{1}{2}$ hadrons when Bjorken-$x$ is finite
($\lambda^{-1/2}\ll x<1$) where supergravity approximation is valid.
Furthermore, they also investigated the case at small-$x$ where the
Pomeron contribution with a trajectory of
$2-\mathcal{O}\left(\frac{1}{\sqrt{\lambda}}\right)$ was found. In
their work, since an infrared cutoff $\Lambda$ is introduced in
order to generate confinement, the model is then called the hard
wall model. There are also some earlier studies\cite{Brower:2002er,
BoschiFilho:2002zs} on high energy scattering in gauge/string
duality. There have been a lot of further developments along this
direction\cite{Brower:2007xg,Brower:2006ea,BallonBayona:2008zi,BallonBayona:2007rs,BallonBayona:2007qr,
Pire:2008zf,Cornalba:2007zb,Cornalba:2008sp,Cornalba:2009ax}. A
saturation picture based on deep inelastic scattering in AdS/CFT is
developed\cite{Hatta:2007he} afterwards and recently reviewed in
Ref.~\cite{Hatta:2008zz}.  In addition, the deep inelastic
scattering off the finite temperature plasma in gauge/string duality
is  studied in Refs.~\cite{Hatta:2007cs,
Mueller:2008bt,Iancu:2009py,Bayona:2009qe,Hassanain:2009xw}.

Recently, the above deep inelastic scattering (DIS) calculation in gauge/gravity duality has been extended to the case of
polarized DIS off the charged dilatino  in Ref.~\cite{Gao:2009ze} and obtained the spin-dependent
structure functions $g_{1}$ and $g_{2}$ for a spin-$\frac{1}{2}$ hadron at finite $x$.
In Ref.~\cite{Hatta:2009ra}, the small $x$ behavior of such spin-dependent
structure functions at large coupling limit was analyzed. Furthermore, the nonforward Compton scattering
has been also investigated in Refs.~\cite{Gao:2009se,Marquet:2010sf}.
Other recent relevant work can be found in Refs.~\cite{Betemps:2010ij,Hatta:2010kt,Cornalba:2010vk}.

In  Ref.~\cite{Polchinski:2002jw} and Ref.~\cite{Gao:2009ze},
both the unpolarized and polarized structure functions when $x$ is finite ($\lambda^{-1/2}\ll x<1$)
are calculated with only minimal interaction. It is found that all the structure functions are
power suppressed and vanish in the large $q^2$ limit. So it is worthwhile to investigate how the structure functions
look  from other possible
interactions. Also, since the minimal interaction  is proportional to
the charge that the dilatino  carries, such interaction does not contribute to the structure functions of the neutral dilatino.
One of our main objects in this paper is to extend the calculation of the structure functions of the charged spin-$\frac{1}{2}$ hadron
---just call it ``proton''---in polarized DIS in Ref.~\cite{Gao:2009ze} to the  neutral spin-$\frac{1}{2}$ hadron
---just call it ``neutron''---through introducing a new interaction --- Pauli interaction
term in $\mbox{AdS}_5$ space. Besides, in Ref.~\cite{Gao:2009ze}, the authors have obtained an
interesting Burkhardt-Cottingham-like sum rule $\int_0^1 dx g_2(x,q^2)=0$, which is completely independent of $\tau$ and $q^2$,
hence another main object of our paper is to investigate whether such sum rule still holds from other possible interaction terms
such as Pauli interaction here.
Since there has been extensive study \cite{deTeramond:2008ht, Brodsky:2008pf,
Brodsky:2007hb, Brodsky:2006uqa, Brodsky:2003px,Grigoryan:2007wn,
Grigoryan:2007my,Grigoryan:2007vg,Abidin:2008hn,Abidin:2008ku,Hong:2007dq, Brodsky:2008pg,
Abidin:2009hr} of the elastic form factors, we will not take this subject into account in our present work.
For simplicity and also consistency with the previous work, we still work in the hard wall model.

This paper is organized as follows. In Sec.\ref{deft}, we recall the
definitions for various structure functions as well as kinematic
variables. In Sec.\ref{hard}, we calculate  the structure functions
of the ``neutron'' at finite $x$ from the Pauli interaction term.
Section \ref{disc} is devoted to the discussions and comments on these
structure functions and compare them with the structure functions of the
charged dilatino which have been calculated in the previous work
where only minimal interaction was considered. We summarize our
results in Sec.\ref{conc}. Finally, in the appendix we will
illustrate how a Pauli interaction term in 5D can arise naturally
from  6D fermion-graviton  coupling  through the usual  Kaluza-Klein
reduction.
\section{Polarized Deep inelastic Scattering}
\label{deft}
\begin{figure}[htbp]
\includegraphics[width=10cm]{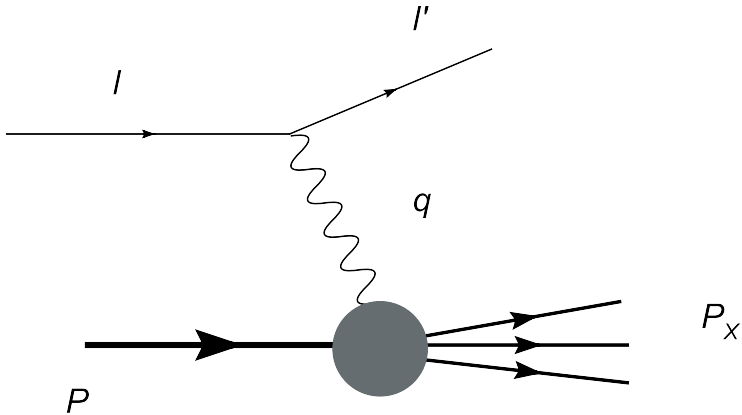}
\caption{The lepton interacts with the
hadron target through the exchange of a virtual photon; the hadron absorbs the virtual photon
and fragments into the final state $X$.}
\label{DIS}
\end{figure}

Deep inelastic scattering  has played an important role in the history
of investigating the internal structure of hadrons. It is the study of lepton-hadron scattering
in the limit that $x$ is fixed, and $q^2\rightarrow \infty$. The basic diagram for such
process is illustrated schematically in Fig.~\ref{DIS}. The structure of the hadron can be
completely characterized by the hadronic tensor $W^{\mu\nu}$, which is defined as
\begin{equation}
  W_{\mu\nu} = \int \! d^4 \xi \, e^{i q{\cdot}\xi} \,
  \langle P,  S |[ J_\mu(\xi), J_\nu(0)] | P,  S \rangle \, ,
  \label{dis3}
\end{equation}
with $J_{\mu}$ being the incident current. In our present work, we will specify the hadron as
the spin-$\frac{1}{2}$ hadron. The hadronic tensor $W_{\mu\nu}$ can be split as
\begin{equation}
  W_{\mu\nu} =
  W_{\mu\nu}^\mathrm{(S)}(q, P) + i \, W_{\mu\nu}^\mathrm{(A)}(q,P,S) \, .
  \label{dis10}
\end{equation}
According to Lorentz and $CP$ invariance, the symmetrical and
antisymmetrical parts can be expressed in terms of 8 independent
structure functions as\cite{Anselmino:1994gn,
Lampe:1998eu}
\begin{eqnarray}
 W_{\mu\nu}^\mathrm{(S)} &=&
 \left( \eta_{\mu\nu}-\frac{q_\mu q_\nu}{q^2}\right) \left[F_1(x,q^2)+\frac{M S\cdot q}{2P\cdot q}g_5(x,q^2)\right]\nonumber\\
& &  - \frac{1}{P{\cdot}q}\left( P_\mu - \frac{P{\cdot}q}{q^2} \, q_\mu \right)\left( P_\nu - \frac{P{\cdot}q}{q^2} \, q_\nu \right)
  \left[F_2(x,q^2)+\frac{M S\cdot q}{P\cdot q}g_4(x,q^2)\right]\nonumber\\
& &  -\frac{M}{2P\cdot q}\left[\left( P_\mu - \frac{P{\cdot}q }{q^2}q_\mu\right)\left(S_\nu - \frac{S{\cdot}q}{P{\cdot}q} \, P_\nu\right)
+\left( P_\nu - \frac{P{\cdot}q }{q^2}q_\nu\right)\left(S_\mu - \frac{S{\cdot}q}{P{\cdot}q} \, P_\mu\right)\right]g_3(x,q^2),\nonumber\\
  W_{\mu\nu}^\mathrm{(A)}
  &=&
  -\frac{M \, \varepsilon_{\mu\nu\rho\sigma} \, q^\rho}{P{\cdot}q}
  \left\{
    S^\sigma \, g_1(x,q^2) +
    \left[ S^\sigma - \frac{S{\cdot}q}{P{\cdot}q} \, P^\sigma \right] g_2(x,q^2)
  \right\} -\frac{\varepsilon_{\mu\nu\rho\sigma}q^\rho P^\sigma}{2P{\cdot}q}
  F_3(x,q^2),
\label{wmunu}
\end{eqnarray}
where $M$ is the mass of the hadron, $S$ is its polarization vector, $q$
is the momentum carried by the current and $P$ is the initial
momentum of the hadron (See Fig.~\ref{DIS}). In deep inelastic
scattering, we define the kinematic variables as the following
\begin{equation}
x=-\frac{q^2}{2P\cdot q} \quad \textrm{and} \quad P_X^2=(P+q)^2.
\end{equation}
The mass of the intermediate state after the scattering is defined
as $M_X^2=s=-P_X^2$. All the structure functions are only functions of
$x$ and $q^2$.

 We will  use the most plus metric throughout this
paper instead of the usual most minus metric in particle physics,
so there are some sign changes in our definitions comparing to the usual definitions in
\cite{Anselmino:1994gn, Lampe:1998eu}.
\section{Polarized structure functions  in the hard wall model}
\label{hard}
According to the conjecture of AdS/CFT, at large 't Hooft parameter, the gauge theories
have a dual string description,  which  can make accurate analytic calculations possible.
For the $3+1$ dimensional conformal gauge theories, the dual string theory lives in the  space $\textrm{AdS}_5\times W$.
The metric in $\textrm{AdS}_5\times W$ space can be written as
\begin{eqnarray}
ds^2=\frac{R^2}{z^2}(\eta_{\mu\nu}dy^\mu dy^\nu+dz^2)+R^2ds_W^2.
\end{eqnarray}
where  $y^\mu$ are identified with the space-time coordinates in the gauge theory and $W$ denotes a five-dimensional compact space.
We will specify such compact space as $S^5$ in our work.
The conformal invariance can be broken through introducing a sharp cut-off $0\leq z \leq z_0\equiv1/\Lambda$,
leading to the mass gap of hadrons. This simple model is the so-called hard wall model.

Following the formalism proposed by Polchinski and
Strassler in Ref.~\cite{Polchinski:2002jw}, the incident current is chosen to be the $\mathcal{R}$-current
which couples to the hadron as an isometry of $S^5$. According to
the AdS/CFT correspondence, the current excites a nonnormalizable
mode of a Kaluza-Klein gauge field at the Minkowski boundary of
the $\textrm{AdS}_5$ space
\begin{equation}
\delta G_{m a} = A_m(y,z) v_a(\Omega) ,
\end{equation}
where $v_a(\Omega)$ denotes a Killing vector on $S^5$ with
$\Omega$ being the angular coordinates on $S^5$. $ A_m(y,z)$ is
the external potential in the gauge theory corresponding to the
operator insertion $n_\mu J^\mu(q)$ on the boundary of the fifth
dimension of the $\mathrm{AdS}_5$ space with the boundary condition
\begin{equation}
\label{boundary-A}
A_\mu(y,0) =  A_\mu(y)|_{\rm 4d}= n_\mu e^{i q \cdot y}.
\end{equation}
This gauge field fluctuation $ A_m(y,r)$ can be viewed as a vector
boson field which couples to the $\mathcal{R}$-current $J^{\mu}$
on the Minkowski boundary, and then propagates into the bulk as a
gravitational wave, and eventually interacts with the supergravity
modes of the dilatino or dilaton.
The gauge field satisfies Maxwell's equation in the bulk, $D_{m}F^{mn}=0$,
which can be explicitly written as
\begin{equation}
\frac{1}{\sqrt{-g}}\partial _{m}\left[ \sqrt{-g}g^{nk}g^{ml}\left( \partial
_{k}A_{l}-\partial _{l}A_{k}\right) \right] =0,
\end{equation}%
where $m$, $n$, ... are indices on $\textrm{AdS}_{5}$. In the Lorentz-like gauge
\[
\partial _{\mu }A^{\mu }+z\partial _{z}\left( \frac{A_{z}}{z}\right) =0,
\]%
the Maxwell equation can be written as
\begin{eqnarray}
-q^{2}A_{\mu }+z\partial _{z}\left( \frac{1}{z}\partial _{z}A_{\mu }\right)
&=&0, \\
-q^{2}A_{z}+\partial _{z}\left( z\partial _{z}\left( \frac{1}{z}A_{z}\right)
\right) &=&0.
\end{eqnarray}%
The solutions to the above equations with the proper boundary conditions $F_{z\mu}(y,z_0)=0$
are given by \footnote{If we  choose  another alternative
gauge invariant boundary conditions $F_{\mu\nu}(y,z_0)=0$,
we will see it  leads to unreasonable  constraint $n^\mu \propto q^\mu$ on the boundary condition (\ref{boundary-A}).}
\begin{eqnarray}
A_{\mu } &=&n_{\mu }e^{iq\cdot y}qz\left[\text{K}_{1}(qz)+c\text{I}_{1}(qz)\right],  \nonumber \\
A_{z} &=&in\cdot qe^{iq\cdot y}z\left[\text{K}_{0}(qz)-c\text{I}_{0}(qz)\right],
\end{eqnarray}
where
\begin{eqnarray}
c={\text{K}_{0}(qz_0)}/{\text{I}_{0}(qz_0)}.
\end{eqnarray}
In the large $q^2$ regime, we can just simply identify $c$ as $0$. However, in the
small $q^2$ regime, such a term contributes as much as the others. This is just the  reason
 the form factor gives rise to logarithmic divergent charge radii for the charged dilatino
in the work \cite{Gao:2009ze} by Gao and Xiao.

Spin-$\frac{1}{2}$ hadrons corresponds to supergravity modes of the
dilatino. In the conformal region the dilatino field can be written as
\begin{equation}
\lambda =\Psi (y,z)\otimes \eta (\Omega )\ ,
\end{equation}%
where $\Psi (y,z)$ is an $SO(4,1)$ spinor on $\mathrm{AdS}_{5}$ and $\eta (\Omega )$
is an $SO(5)$ spinor on $S^{5}$. The wave-function $\Psi $ satisfies a
five-dimensional Dirac equation in $\mathrm{AdS}_{5}$ space. Let us first review how to derive this
five-dimensional Dirac equation in the following.

A convenient choice of vielbein is given by
\begin{equation}
e_{m}^{a}=\frac{R}{z}\delta _{m}^{a},\ \ e^{ma}=\frac{z}{R}\eta ^{ma},\ \
e_{a}^{m}=\frac{z}{R}\delta _{a}^{m}
\end{equation}%
where $m=0,1,2,3,5$. The Levi-Civita connection is given by
\begin{equation}
\Gamma _{mn}^{p}=\frac{1}{2}g^{pq}(\partial _{n}g_{mq}+\partial
_{m}g_{nq}-\partial _{q}g_{mn})
\end{equation}%
Here we use $a,b,c$ to denote indices in flat space, and $m,n,p,q$ to denote
indices in curved space ($\mathrm{AdS}_{5}$ space). In addition, the Greek indices $%
\mu ,\nu $ are defined in Minkowski space. From the metric, one knows
\begin{equation}
g_{mn}=\frac{R^{2}}{z^{2}}\eta _{mn}.
\end{equation}%
It is straightforward to work out the Levi-Civita connection in $\mathrm{AdS}_{5}$
space
\begin{equation}
\Gamma _{\mu \nu }^{5}=\frac{1}{z}\eta _{\mu \nu },\ \ \Gamma _{55}^{5}=-%
\frac{1}{z},\ \ \Gamma _{\nu 5}^{\mu }=-\frac{1}{z}\delta _{\nu }^{\mu }
\end{equation}%
From vielbein and Levi-Civita connection, we can have the spin connection
\begin{equation}
\omega _{m}^{ab}=e_{n}^{a}\partial _{m}e^{nb}+e_{n}^{a}e^{pb}\Gamma
_{pm}^{n}.
\end{equation}%
The only nonvanishing spin connections are%
\begin{equation}
\omega _{\mu }^{5\nu }=-\omega _{\mu }^{\nu 5}=\frac{1}{z}\delta _{\mu
}^{\nu }.
\end{equation}%
Using above results, the operator $D\hspace{-8pt}\slash$ can be cast into
\begin{equation}
D\hspace{-8pt}\slash=g^{mn}e_{n}^{a}\gamma _{a}\left( \partial _{m}+\frac{1}{%
2}\omega _{m}^{bc}\Sigma _{bc}\right) =\frac{z}{R}\left( \gamma ^{5}\partial
_{z}+\gamma ^{\mu }\partial _{\mu }-\frac{2}{z}\gamma ^{5}\right) ,
\end{equation}%
with $\Sigma _{\mu 5}=\frac{1}{4}\left[ \gamma _{\mu },\gamma _{5}\right] $.
The free dilatino field in $\mathrm{AdS}_{5}$ space satisfies the Dirac equation
\begin{equation}
(D\hspace{-8pt}\slash-m)\Psi =\frac{z}{R}\left( \gamma ^{5}\partial
_{z}+\gamma ^{\mu }\partial _{\mu }-\frac{2}{z}\gamma ^{5}-\frac{mR}{z}%
\right) \Psi =0.
\end{equation}
Its normalizable solution is given by \cite{Mueck:1998iz},
\begin{equation}
\Psi (z,y)=Ce^{ip\cdot y}z^{\frac{5}{2}}\left[
J_{mR-1/2}(Mz)P_{+}+J_{mR+1/2}(Mz)P_{-}\right] u_{\sigma }
\end{equation}
where
\begin{equation}
p\hspace{-4.5pt}\slash u_{\sigma }=-iMu_{\sigma }(\sigma =1,2),\ \
M^{2}=-p^{2},\ \ P_{\pm }=\frac{1}{2}(1\pm \gamma ^{5})
\end{equation}

For the initial hadron, by assuming $Mz \ll 1$ in the interaction region and expanding the Bessel functions up to linear term in $M$, one gets
\begin{equation}
\psi_{\rm i} \approx e^{i P \cdot y} \frac{c'_{\rm i}z_0^{3/2}}{ R^{9/2} } (\frac{z}{z_0})^{ mR + 2}
\left[P_+u_{\rm i\sigma}+\frac{M z}{2(mR+1/2)} P_- u_{\rm i\sigma}\right].
\end{equation}
\begin{equation}
\bar{\psi}_{\rm i} \approx e^{-i P \cdot y} \frac{c'_{\rm i}z_0^{3/2}}{ R^{9/2} }
(\frac{z}{z_0})^{ mR + 2} \left[ \bar{u}_{\rm i\sigma}P_- +\frac{M z}{2(mR+1/2)}  \bar{u}_{\rm i\sigma}P_+\right].
\end{equation}
For the intermediate hadron, $M_X \gg 1/z_0$ and
\begin{equation}
\psi_{X} \approx e^{i (P+q) \cdot y} \frac{c'_{X} M_X ^{1/2}
z^{5/2}}{R^{9/2} z_0^{1/2}}
\Bigl[
J_{mR - 1/2}(M_X z) P_+ + J_{mR + 1/2}(M_X z) P_- \Bigr]
u_{X\sigma} \ .
\end{equation}
\begin{equation}
\bar{\psi}_{X} \approx e^{-i (P+q) \cdot y} \frac{c'_{X} M_X ^{1/2}
z^{5/2}}{R^{9/2}z_0^{1/2}}
\bar{u}_{X\sigma}\Bigl[
P_- J_{mR - 1/2}(M_X z)  + P_+ J_{mR + 1/2}(M_X z)  \Bigr]
 \ .
\end{equation}

In Ref.~\cite{Polchinski:2002jw} and Ref.~\cite{Gao:2009ze},
the interaction between the Kaluza-Klein gauge field and charged dilatino is given by
the minimal coupling
\begin{eqnarray}
S_{int}^{M} &=&iR^5\int d^{5}x\sqrt{-g}\,{\mathcal{Q}}A_{m}%
{e^{m}}_{a}\overline{\Psi }\gamma ^{a}\Psi,
\end{eqnarray}%
However, since we are now interested in the effect of the other possible interactions and
the  neutral dilatino in such ${\cal R}$ current, where the above minimal interaction
will not contribute, we need introduce a new interaction
term ---the Pauli interaction term such that \cite{Abidin:2009hr}

\begin{eqnarray}
S_{int}^P
&=& \kappa R^6\int d^{5}x\sqrt{-g}\,F_{m n}%
{e^{m}}_{a}{e^{n}}_{b}\overline{\Psi }\ [\gamma ^{a},\gamma ^{b}]\Psi .
\end{eqnarray}%
Actually such term can be derived very naturally from  Kaluza-Klein reduction of higher dimensional
fermion-graviton  coupling. We put such derivation into the final appendix in our present work.

With the Pauli interaction action at hand, following the same line in Ref.~\cite{Gao:2009ze},
we can compute the matrix element
\begin{eqnarray}
{\cal M}^\mu&=& \langle {P_X ,\sigma'}|  J^\mu(0) |{P,\sigma}\rangle\nonumber\\
&=&\ \ \ \frac{1}{2\pi}\left( M_X/z_0\right)^{1/2}\left\{c_1 q\overline{u}_{f\sigma^{\prime }} [\,q\hspace{-6pt}\slash, \gamma^\mu]P_{-}u_{i\sigma }
+c_2 q\overline{u}_{f\sigma^{\prime }} [\,q\hspace{-6pt}\slash, \gamma^\mu]P_{+}u_{i\sigma }\right.\nonumber\\
& &\left.+c_3 \overline{u}_{f\sigma^{\prime }} (q^\mu q\hspace{-6pt}\slash-q^2\gamma^\mu) P_{-}u_{i\sigma }
+c_4 \overline{u}_{f\sigma^{\prime }} (q^\mu q\hspace{-6pt}\slash-q^2\gamma^\mu) P_{+}u_{i\sigma }\right\},
\end{eqnarray}
and  its complex conjugate
\begin{eqnarray}
{\cal M}^{*\mu}&=&\langle {P,\sigma}|  J^\mu(0) |{P_X ,\sigma'}\rangle\nonumber\\
&=&\ \ \ \frac{1}{2\pi}\left(M_X/z_0\right)^{1/2}\left\{-c_1 q\overline{u}_{i\sigma}
[\,q\hspace{-6pt}\slash, \gamma^\mu]P_{+}u_{f\sigma^{\prime } }
-c_2 q\overline{u}_{i\sigma} [\,q\hspace{-6pt}\slash, \gamma^\mu]P_{-}u_{f\sigma^{\prime } }\right.\nonumber\\
& &\left.+c_3 \overline{u}_{i\sigma} (q^\mu
q\hspace{-6pt}\slash-q^2\gamma^\mu) P_{-}u_{f\sigma^{\prime } }
+c_4 \overline{u}_{i\sigma} (q^\mu
q\hspace{-6pt}\slash-q^2\gamma^\mu) P_{+}u_{f\sigma^{\prime } }\right\}.
\end{eqnarray}
The coefficients $c_i (i=1,2,3,4)$ are given by

\begin{eqnarray}
c_1&=&\frac{c_0 M}{2(\tau-1)}\int d z z^{\tau+2} \, K_1(qz) J_{\tau-2}(M_X z)\nonumber\\
&=&c_0  2^\tau M_X^{\tau-2}(M_X^2+q^2)^{-\tau-2}M q\tau\left[q^2(\tau-1)-2M_X^2\right]\Gamma(\tau-1),\\
c_2&=&c_0\int d z z^{\tau+1} \, K_1(qz) J_{\tau-1}(M_X z)\nonumber\\
&=&c_0 2^\tau M_X^{\tau-1}(M_X^2+q^2)^{-\tau-1}q \Gamma(\tau+1),\\
c_3&=&\frac{c_0 M}{2(\tau-1)}\int d z z^{\tau+2} \, K_0(qz) J_{\tau-1}(M_X z)\nonumber\\
&=&c_0  2^\tau M_X^{\tau-1}(M_X^2+q^2)^{-\tau-2}M\tau\left[\tau q^2-M_X^2\right]\Gamma(\tau-1),\\
c_4&=&c_0 \int d z z^{\tau+1} \, K_0(qz) J_{\tau-2}(M_X z)\nonumber\\
&=&c_0 2^\tau M_X^{\tau-2}(M_X^2+q^2)^{-\tau-1}\left[q^2(\tau-1)-M_X^2\right]\Gamma(\tau),
\end{eqnarray}
where $\tau\equiv mR+\frac{3}{2}$
and $c_0 =2\pi c_i' c_X'z_0^{-\tau+1}$.
 In the above calculation, we have relaxed the upper
limit of integration from $z_0=1/\Lambda$ to $\infty$. Now Let us
continue to calculate the hadronic tensor

\begin{eqnarray}
W_{\mu \nu}&=&W_{\mu\nu}^{(S)}+iW_{\mu\nu}^{(A)}
=(2\pi)^3\sum_{X} \delta \left(M_X^2+(P+q)^2\right){\cal M}_\mu {\cal M}_{\nu}^*
\end{eqnarray}
In  large $q^2$ limit, we can make the approximation
\begin{equation}
\sum_{X}\delta \left(M_X^2+(p+q)^2\right)\simeq \frac{1}{2\pi M_X \Lambda}.
\end{equation}
Together with $M_X^2+q^2=q^2/x$ and $M_X=q\sqrt{(1-x)/{x}}$,
we can write the  symmetric $W_{\mu\nu}^{(S)}$ and antisymmetric $W_{\mu\nu}^{(A)}$ as,
respectively,

\begin{eqnarray}
W_{\mu\nu}^{(S)}&=&\frac{4q^6}{x}\left[2c_2\left(\frac{1-x}{x}\right)^{1/2}+ c_4 \right]^2
\left[1+\frac{2M(S\cdot q)}{2 P\cdot q}\right]\left(\eta_{\mu\nu}-\frac{q_\mu q_\nu}{q^2}\right)\nonumber\\
& &-\frac{8q^6}{x }(4c_2^2+c_4^2)\left[1+\frac{M(S\cdot q)}{P\cdot q}\right]
\frac{1}{P\cdot q}\left(P_\mu-\frac{P\cdot q}{q^2}q_\mu\right)\left(P_\nu-\frac{P\cdot q}{q^2}q_\nu\right)\nonumber\\
& &-\frac{8q^6}{x}\left[(4c_2^2+c_4^2)+(c_2 c_3-c_1 c_4)\frac{2q}{x M}\right]\frac{M}{2P\cdot q}\nonumber\\
& &\times\left[\left(P_\mu-\frac{P\cdot q}{q^2}q_\mu\right)\left(S_\nu-\frac{S\cdot q}{P\cdot q}P_\nu\right)
+\left(P_\nu-\frac{P\cdot q}{q^2}q_\nu\right)\left(S_\mu-\frac{S\cdot q}{P\cdot q}P_\mu\right)\right]
\end{eqnarray}

\begin{eqnarray}
W_{\mu\nu}^{(A)}&=& -\frac{8q^6}{x}\left[2c_2\left(\frac{1-x}{x}\right)^{1/2}+ c_4\right]^2
\frac{\epsilon_{\mu\nu\alpha\beta}q^\alpha P^\beta}{2P\cdot q}
-\frac{4q^6}{x}\left[2c_2\left(\frac{1-x}{x}\right)^{1/2}+ c_4\right]^2\frac{M\epsilon_{\mu\nu\alpha\beta}q^\alpha S^\beta}{P\cdot q}
\nonumber\\
& &+\frac{2q^6}{x^2}\left[(4c_2^2+c_4^2)+2(4c_1 c_2-c_3 c_4)\frac{q}{M} \left(\frac{1-x}{x}\right)^{1/2}
+2(c_1 c_4+c_2 c_3)\frac{q}{M}\frac{2x-1}{x} \right]\nonumber\\
& &\hspace{1cm}\times\frac{M\epsilon_{\mu\nu\alpha\beta}q^\alpha}{P\cdot q} \left(S^\beta-\frac{S\cdot q}{P\cdot q}P^\beta\right)
\end{eqnarray}
To obtain the above final results,  we have used the identity

\begin{eqnarray}
\epsilon^{\mu\nu\alpha\beta}q_\alpha\left[(q\cdot S)P_\beta-(P\cdot q)S_\beta \right]= q^\mu\epsilon^{\nu\alpha\beta\gamma}P_\alpha q_\beta S_\gamma
    -q^\nu\epsilon^{\mu\alpha\beta\gamma}P_\alpha q_\beta S_\gamma
    -q^2\epsilon^{\mu\nu\alpha\beta}P_\alpha S_\beta
\end{eqnarray}
Comparing with Eq.~(\ref{wmunu}), we can obtain all the structure functions of the ``neutron,''

\begin{eqnarray}
F_1^\mathrm{n}&=&g_1^\mathrm{n}=\frac{F_3^\mathrm{n}}{2}=\frac{g_5^\mathrm{n}}{2}\nonumber\\
&=&16\pi \kappa^2 A'\left({\Lambda^2}/{q^2}\right)^{\tau-1} x^{\tau+1}(1-x)^{\tau-2}\left[1-\tau(2-x)\right]^2,\\
F_2^\mathrm{n}&=&g_4^\mathrm{n}=32\pi\kappa^2 A'\left({\Lambda^2}/{q^2}\right)^{\tau-1} x^{\tau+1}(1-x)^{\tau-2}\left[1-\tau x(2-4\tau +3\tau x)\right],\\
g_2^\mathrm{n}&=&-8\pi \kappa^2 A'\left({\Lambda^2}/{q^2}\right)^{\tau-1}x^{\tau}(1-x)^{\tau-2}\frac{1}{\tau-1}\nonumber\\
& &\times\left[\tau(2\tau-5)+2\tau x(\tau^2-10\tau +8)+\tau x^2(7\tau^2+17\tau-6)-6\tau^2 x^3(\tau+1)-1\right],\\
g_3^\mathrm{n}&=&32\pi \kappa^2 A'\left({\Lambda^2}/{q^2}\right)^{\tau-1}x^{\tau+1}(1-x)^{\tau-2}\frac{1}{\tau-1}\nonumber\\
& &\times\left[\tau(4x-3)-\tau^2(x^2+7\tau x^2-4x-6\tau x+2)-1\right].
\end{eqnarray}
It is  straightforward to compute the moments of all the structure functions
when the contributions from $x\ll \lambda^{-1/2}$ are negligible.
Typically there are four different kinds of moments, e.g.,
\begin{eqnarray}
\label{g1nn}
\int_{0}^{1} g_1^\mathrm{n}\left(x, q^2\right) x^{n-1}\textrm{d}x
&=& 16\pi\kappa^2 A^{\prime } (\Lambda^2/q^2)^{\tau-1}
\left[n^2(\tau-1)+n(\tau-1)(6\tau+1)\right.\nonumber\\
& &\left.+\tau(9\tau^2-4\tau-4)\right]\frac{\Gamma(\tau)\Gamma(\tau+n+1)}{\Gamma(2\tau+n+2)}\\
\label{g2nn}
\int_{0}^{1} g_2^\mathrm{n}\left(x, q^2\right) x^{n-1}\textrm{d}x
&=&-8\pi\kappa^2 A^{\prime } (\Lambda^2/q^2)^{\tau-1}
\left[n^3(3\tau^2-4\tau+1)+2n^2\tau(10\tau^2-10\tau+1)\right.\nonumber\\
& &\left.+n(33\tau^4-31\tau^3-9\tau^2-1)+\tau(16\tau^4-13\tau^3-13\tau^2-6\tau-2)\right]\nonumber\\
& &\times\frac{\Gamma(\tau)\Gamma(n+\tau)}{(\tau-1)\Gamma(2\tau+n+2)}\\
\label{g3nn}
\int_{0}^{1} g_3^\mathrm{n}\left(x, q^2\right) x^{n-1}\textrm{d}x
&=&32\pi\kappa^2 A^{\prime } (\Lambda^2/q^2)^{\tau-1}
\left[n^2(\tau^2-1)+n(3\tau^3+3\tau^2-1)+2\tau(\tau+1)^3\right]\nonumber\\
& &\times\frac{\Gamma(\tau)\Gamma(n+\tau+1)}{(\tau-1)\Gamma(2\tau+n+2)}\\
\label{F2nn}
\int_{0}^{1} F_2^\mathrm{n}\left(x, q^2\right) x^{n-1}\textrm{d}x
&=&32\pi\kappa^2 A^{\prime } (\Lambda^2/q^2)^{\tau-1}
\left[n^2(\tau^2-1)+n(6\tau^2-\tau-1)+5\tau^3\right]\nonumber\\
& &\times\frac{\Gamma(\tau)\Gamma(n+\tau+1)}{\Gamma(2\tau+n+2)}
\end{eqnarray}
In order to discuss these results further and make comparisons with
the previous work by Gao and Xiao in Ref.~\cite{Gao:2009ze}, it is
helpful to rewrite the structure functions of the ``proton'' that have
been obtained in Ref.~\cite{Gao:2009ze},
\begin{eqnarray}
 F_{1}^\mathrm{p} &=&\frac{F_{2}^\mathrm{p}}{2}=\frac{F_{3}^\mathrm{p}}{2}=g_{1}^\mathrm{p}
 =\frac{g_{3}^\mathrm{p}}{2}=\frac{g_{4}^\mathrm{p}}{2}=\frac{g_{5}^\mathrm{p}}{2}
 =\frac{\pi}{2} A^{\prime } {\cal{Q}}^2(\Lambda^2/q^2)^{\tau-1}x^{\tau+1}(1-x)^{\tau-2},\\
 g_2^\mathrm{p}&=&\left(\frac{1}{2}\frac{\tau +1}{\tau-1}-\frac{x\tau
}{\tau-1}\right)\frac{\pi}{2} A^{\prime }
{\cal{Q}}^2(\Lambda^2/q^2)^{\tau-1}x^{\tau}(1-x)^{\tau-2}.
\end{eqnarray}
Typically there are just two different kinds of moments, e.g.,
\begin{eqnarray}
\label{g1pn}
\int_{0}^{1} 2g_1^\mathrm{p}\left(x, q^2\right) x^{n-1}\textrm{d}x&=& \pi A^{\prime } {\cal{Q}}^2(\Lambda^2/q^2)^{\tau-1} \frac{\Gamma\left(\tau-1\right)\Gamma\left(\tau+n+1\right)}{\Gamma\left(2\tau+n\right)} \\
\label{g2pn}
\int_{0}^{1} 2g_2^\mathrm{p}\left(x, q^2\right) x^{n-1}\textrm{d}x&=& \pi A^{\prime } {\cal{Q}}^2(\Lambda^2/q^2)^{\tau-1} \frac{\Gamma\left(\tau-1\right)\Gamma\left(\tau+n\right)}{\Gamma\left(2\tau+n\right)} \frac{1-n}{2}.
\end{eqnarray}
In all the above expressions, we have defined $A'=\pi c_i^2 c_X^2 2^{2\tau} \Gamma^2(\tau)$.

\section{Discussions}
\label{disc}

\begin{figure}[tbp]
\begin{center}
\includegraphics[width=12cm]{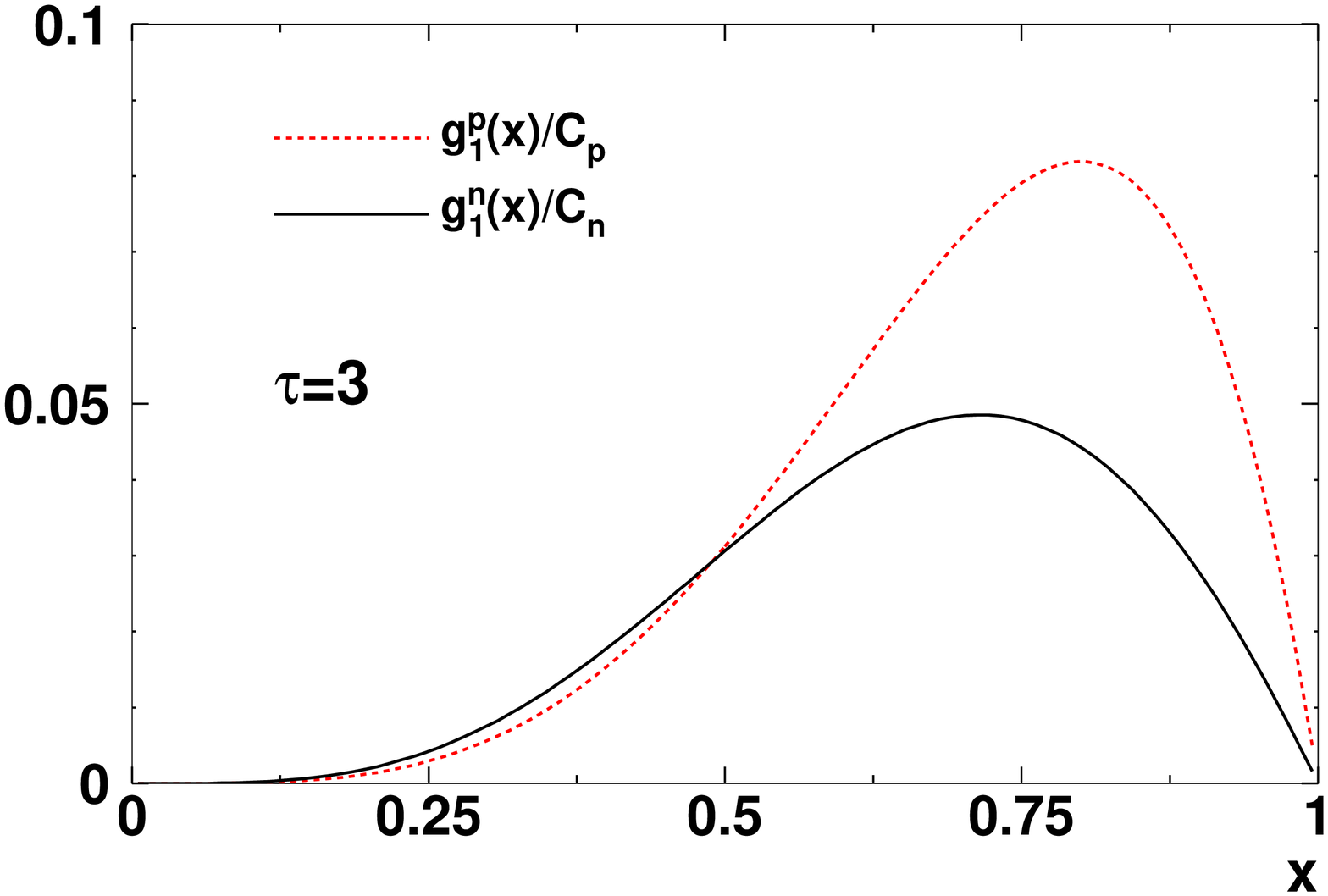}
\end{center}
\caption[*]{Comparison between  the structure functions of $g_1^p$ and $g_1^n$.}
\label{g1}
\end{figure}

\begin{figure}[tbp]
\begin{center}
\includegraphics[width=12cm]{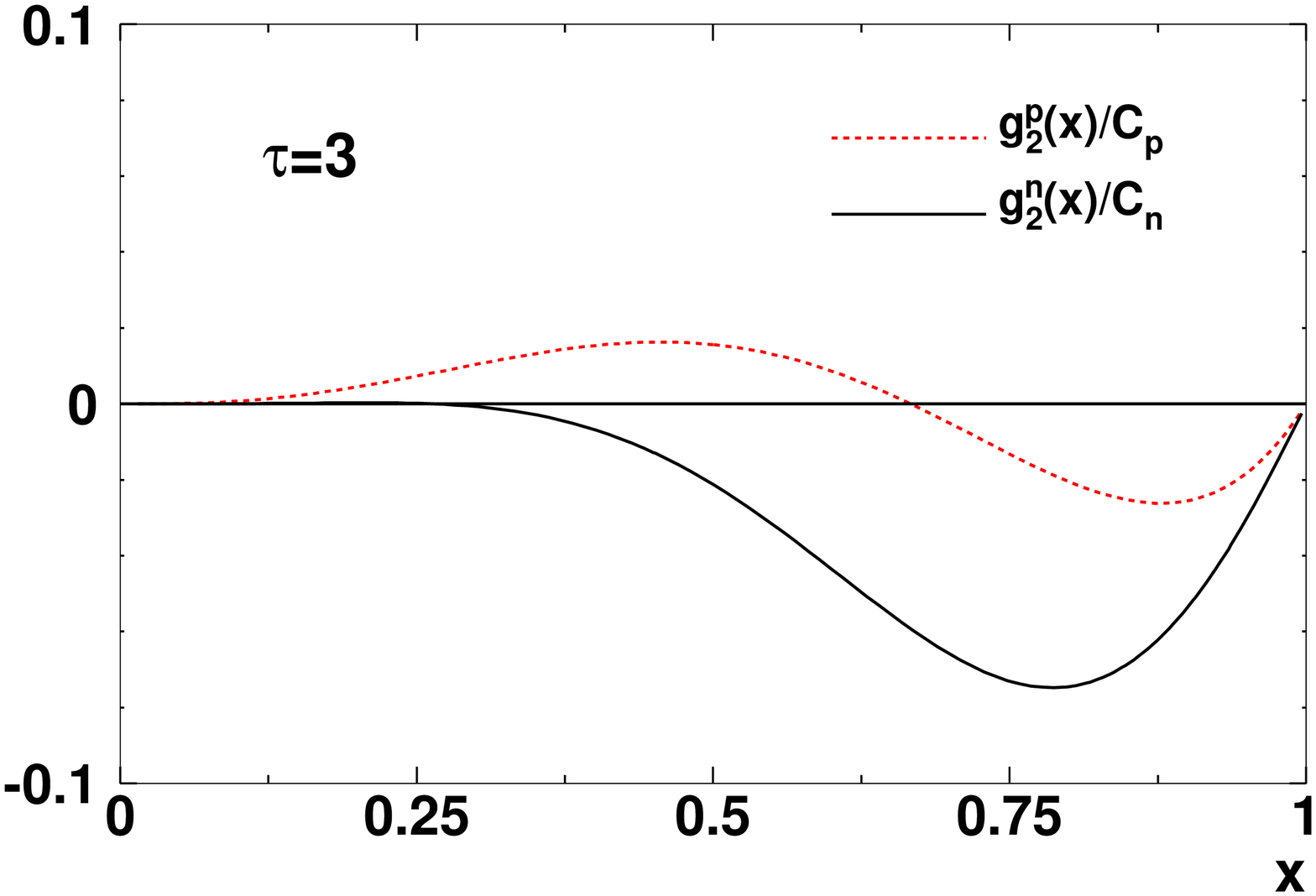}
\end{center}
\caption[*]{Comparison between  the structure functions of  $g_2^p$ and  $g_2^n$.}
\label{g2}
\end{figure}

\begin{figure}[tbp]
\begin{center}
\includegraphics[width=12cm]{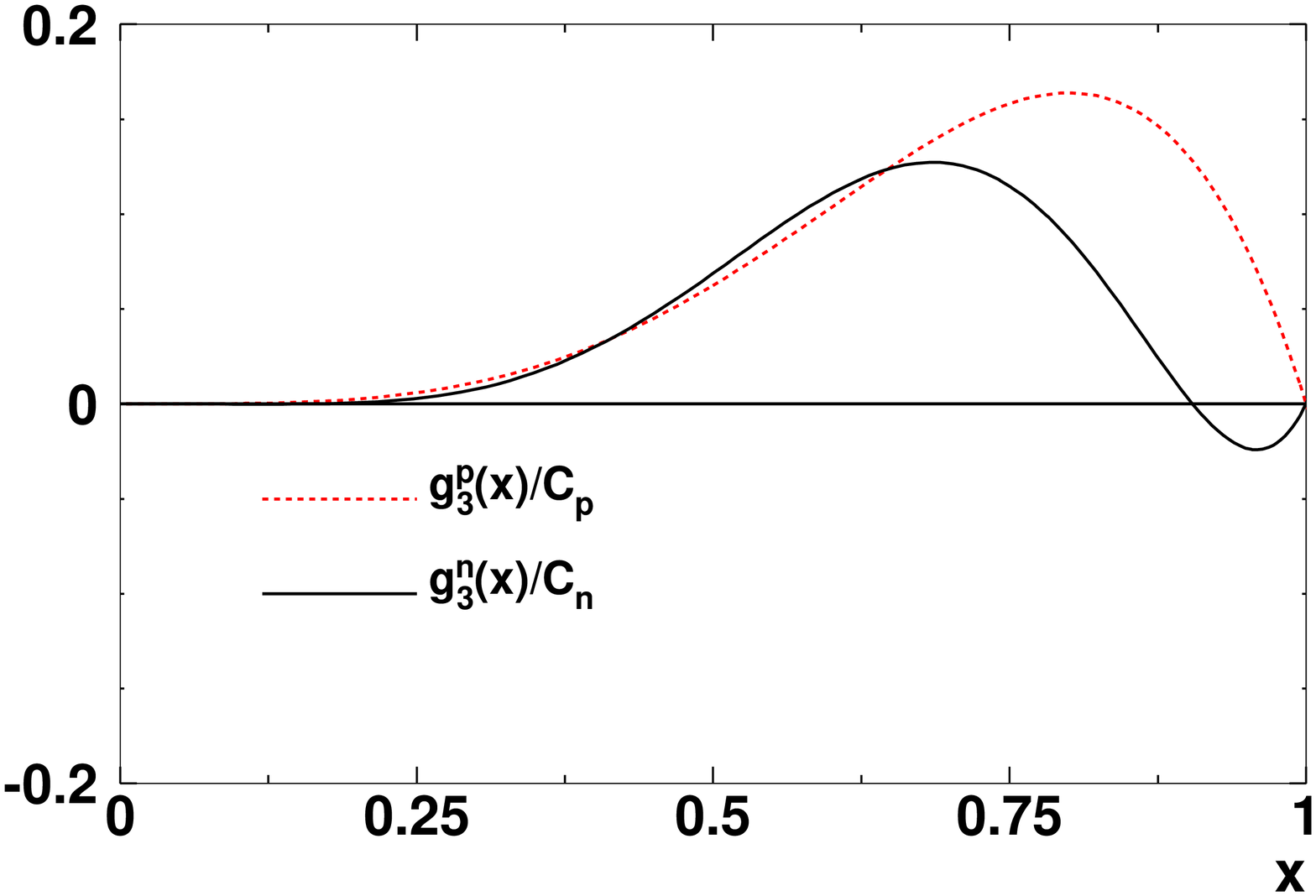}
\end{center}
\caption[*]{Comparison between  the structure functions of $g_3^p$ and $g_3^n$.}
\label{g3}
\end{figure}

\begin{figure}[tbp]
\begin{center}
\includegraphics[width=12cm]{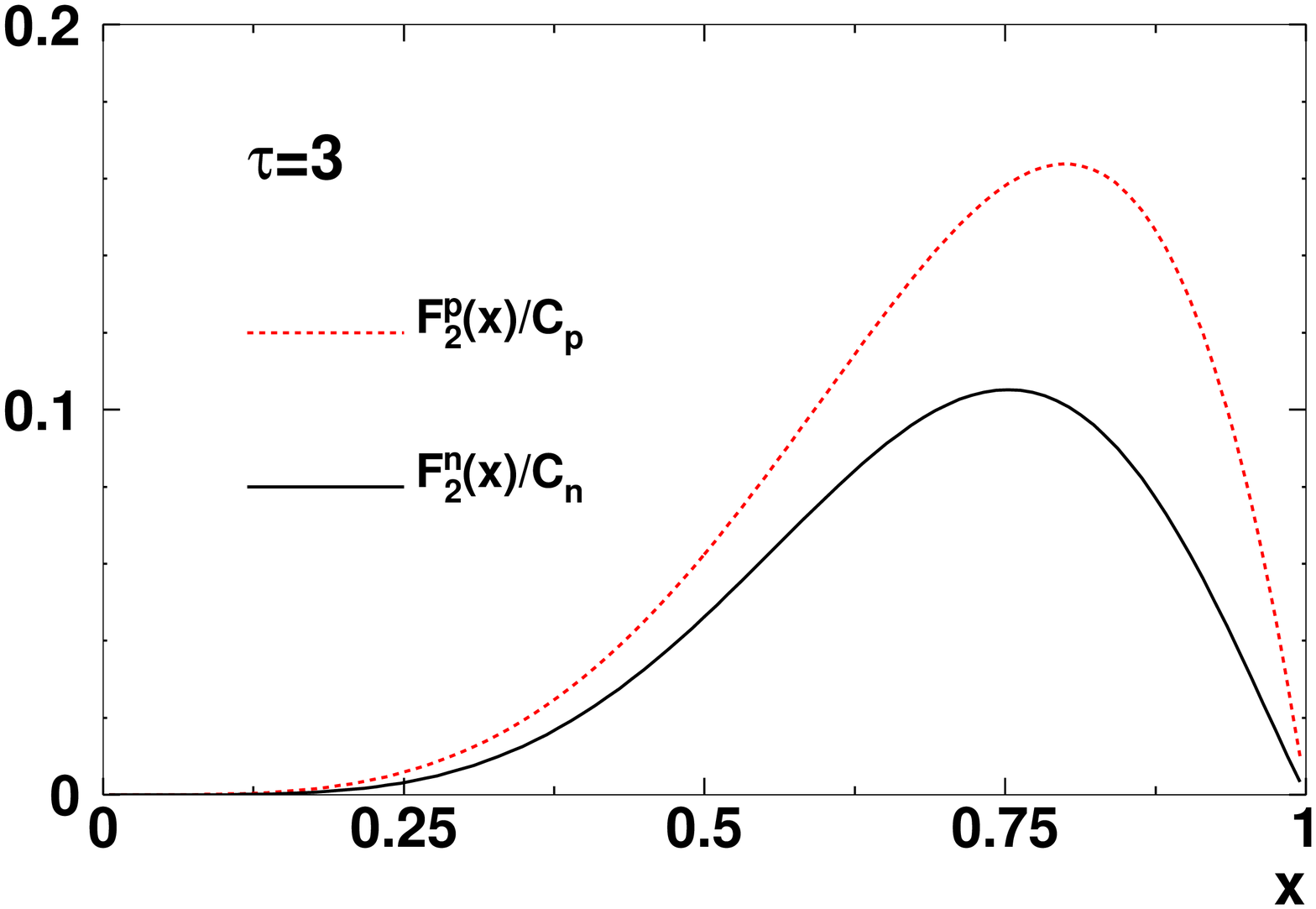}
\end{center}
\caption[*]{Comparison between  the structure functions of $F_2^p$ and $F_2^n$.}
\label{f2}
\end{figure}

In this section, we focus on the interpretation of the structure
functions of the ``neutron'' from Pauli interaction that we obtained from last section using gauge/string
duality and compare these results with the ones of the ``proton'' in Ref.~\cite{Gao:2009ze}, where only minimal interaction was included.

\begin{itemize}
\item Just like what we did in Ref.~\cite{Gao:2009ze},
 only the linear term in $M$ is kept in the initial wavefunction and throughout the calculation,
the results shown above, whether the ``neutron'' or the ``proton,'' are from leading order calculations.
The corrections are of order $\frac{M^2}{q^2}$ or $\frac{\Lambda^2}{q^2}$.
\item The power order $\Lambda/q$ of the structure functions of the ``neutron'' from  only the Pauli interaction are
the same as that  from only minimal interaction.
Ignoring the relative magnitude of $\kappa$ and ${\cal Q}$, but only from the naive dimensional analysis, one might expect that the Pauli interaction
will lead to less power order $\Lambda/q$ by one than that from the minimal interaction.
However, when the warp factor $e_{a}^{m}$ is taken into account, the extra $q$ will combine with $z$ from $e_{a}^{m}$
and lead to $q z_{int}\sim 1$, which results in the same power suppression as the minimal interaction.
\item  The relations $F_1^\mathrm{n} =\frac{F_3^\mathrm{n}}{2}=\frac{g_5^\mathrm{n}}{2}$ and $F_2^\mathrm{n} =g_4^\mathrm{n}$
still hold from the Pauli interaction, but the relation $F_1^\mathrm{p}=\frac{F_2^\mathrm{p}}{2}=\frac{g_3^\mathrm{p}}{2}$
from the minimal interaction is broken. The differences of the dependence on $x$
of all the structure functions are illustrated in Figs.\ref{g1}--\ref{f2},
where the  coefficient  $C_\mathrm{p}=\frac{1}{2}\pi A^{\prime } {\cal{Q}}^2(\Lambda^2/q^2)^{\tau-1}$ and
$C_\mathrm{n}=200 \pi A^{\prime } {\kappa}^2(\Lambda^2/q^2)^{\tau-1}$.
\item  In QCD, there is an interesting inequality $F_1 \geq g_1$\cite{Manohar:1992tz}. In Ref.\cite{Gao:2009ze}, we have found
that $F_1 =g_1$, i.e. the bound is saturated. Here we see that the saturation condition $F_1 =g_1$ still holds,
which indicates that initial hadron is completely polarized. This implies that the struck dilatino just tunnels or shrinks to smaller size of order the inverse momentum transfer during the scattering. As a result, the structure function exhibits a power law behavior in terms of the $q^2$ dependence which comes from the tunneling probability\cite{Polchinski:2001tt, Polchinski:2002jw}.
\item For all the moments Eq.~(\ref{g1nn})--Eq.~(\ref{F2nn}), Eq.~(\ref{g1pn}) and Eq.~(\ref{g2nn}),
we expect that the moments are correct at least for $n >2$ where the low-x contributions are negligible.
When one sets $n=1$ for $g_2^\mathrm{p}$, there is an interesting sum rule
\begin{equation}
\int_0^1 \textrm{d}x g_2^\mathrm{p}\left(x, q^2\right) =0,
\end{equation}
which is completely independent of $\tau$ and $q^2$. In QCD, this sum rule is known as
the Burkhardt-Cottingham sum rule\cite{Burkhardt:1970ti}in large $Q^2$ limit. However,
 this sum rule can be invalidated by non-Regge divergence at low-$x$. Now let us set
 $n=1$ for $g_2^\mathrm{n}$,we can have
\begin{equation}
\int_0^1 \textrm{d}x g_2^\mathrm{n}\left(x, q^2\right) =
-32\pi \kappa^2 A'\left({\Lambda^2}/{q^2}\right)^{\tau-1}
\left(4\tau^3+\tau^2-7\tau-1\right)\frac{\Gamma(\tau+1)\Gamma(\tau+2)}{(\tau-1)\Gamma(2\tau+3)}.
\end{equation}
It is obvious that such sum rule which holds for minimal interaction
in the classic supergravity approximation is broken due to
introducing the Pauli interaction term.
In this place, it is a good opportunity to compare the above conclusion with  QED  but with an extra
nonrenormalizable Pauli interaction term introduced, i.e.
\begin{eqnarray}
S^{QED}_{int}=\int d^4 y \left(i{\cal{Q}}\bar\psi A\hspace{-7pt}\slash\psi +\kappa F_{\mu \nu}
\bar{\psi }\ [\gamma ^{\mu},\gamma ^{\nu}]\psi\right).
\end{eqnarray}
where we just specify $\psi$ as a quark field
\footnote{In the realistic QED, from the viewpoint of effective field theory, the Pauli interaction contribution
is suppressed by $q/M_P$, where $q$ is the energy scale in which we are working and $M_P$ is Planck energy scale.
Hence such contribution is highly suppressed when $q$ is in the scale of GeV or TeV. Since we just want to
show the pure effect of such Pauli interaction term, we will neglect such realistic issues. }.
 It is easy to verify that,
in the tree diagram level of ${\cal Q}$ or $\kappa$ and twist-3 level of $m/q$
($m$ denotes the mass of the quark),
the pure minimal interaction  results in
\begin{eqnarray}
F_1(x,q^2)&=&2F_2(x,q^2)=a_m^F{\cal Q}^2\delta(x-1),\\
g_1(x,q^2)&=&a_m^g{\cal Q}^2\delta(x-1),\ \  g_2(x,q^2)=0,
\end{eqnarray}
 while the pure Pauli interaction results in
\begin{eqnarray}
& &F_1(x,q^2)=a_P^F{\kappa}^2\delta(x-1),\ \ F_2(x,q^2)=0,\\
& &g_1(x,q^2)=0, \ \ g_2(x,q^2)=-a_P^g{\kappa}^2\delta(x-1),
\end{eqnarray}
where $a_m^F$, $a_m^g$, $a_P^F$, and $a_P^g$ are all positive coefficients, which is irrelevant with our current problem.
Hence, it is obvious that,  similar to the $\mathrm{AdS}_{5}$ space, the Pauli interaction term always makes the
Burkhardt-Cottingham sum rule invalidated. It should be clarified that the effective Pauli interaction
can be produced from high order contributions in usual QED
or  instantons \cite{Kochelev:2003cp} in usual QCD without the Pauli interaction term
in the original Lagrangian. Actually, from the spirit of the conjecture of AdS/CFT, the Pauli interaction introduced in our present work
in $\mathrm{AdS}$ space is equivalent to  summing over all the loop contributions in CFT side.

\item  It is obvious that  the moments of all the structure functions
are power suppressed, for sufficiently large $q^2\rightarrow \infty$, all these integrals vanishes.
Actually in all our calculations, the results  are only valid at large t' Hooft coupling $\lambda$
and finite $x$ with $\lambda^{-1/2}\ll x<1$. In order to  take into account the moments of the
structure functions completely, we need to consider the very small $x$ case.
For example, the missing contributions from the Pomeron exchanges  to $F_1$ and $F_2$  peaks around $x=0$.
\begin{equation}
x F_1 \sim F_2 \propto x^{-1+\mathcal{O}(1/\sqrt{\lambda})}
\end{equation}
where the correction to the Pomeron intercept arises from the
curvature of $\mathrm{AdS}_5$. Such Pomeron contribution will
survive in the large $q^2$ limit and give us a nonvanishing second
moment of $F_1$\cite{Polchinski:2002jw, Hatta:2007he}, which make
energy momentum conserved. There is a similar contribution to
$g_1$ at small-$x$ which yields a singular\cite{Hatta:2009ra}
\begin{equation}
g_1\sim \frac{1}{x^{\alpha_{R1}}},
\end{equation}
with $\alpha_{R1} =1-\mathcal{O}(\frac{1}{\sqrt{\lambda}})$ when $x$
is extremely small. This contribution will also survive in large
$q^2$ limit and yield a finite first moment. This may indicate that
most of the hadron spin is carried by the small-$x$ constituents
inside the hadron. The detailed discussions on the small-$x$ limit of
the $g_1$ structure function can be available in
Ref.~\cite{Hatta:2009ra}.
\item Just repeat the arguments in Ref.~\cite{Gao:2009ze} on the  parity violating structure functions $F_3$, $g_3$, $g_4$, and $g_5$.
These parity violating structure functions are as large as the $F_2$
structure function due to the reason that the dilatino is
right-handed fermion in massless limit. They are tightly related to
the peculiar wavefunction of the dilatino. However, we expect that
$g_1$ and $g_2$ may exhibit some common features of the polarized
structure functions of spin-$\frac{1}{2}$ hadrons in the
nonperturbative region when the coupling is large.
\item Phenomenologically, we can just regard $\kappa$ as a free parameter, which can
be fixed by the experimental values of the ``neutron'' magnetic moments. The
detailed discussion and fitting results can be found in Ref.\cite{Abidin:2009hr}.
\end{itemize}

\section{Conclusion}
\label{conc}
Through introducing the Pauli interaction term in the action in the $\textrm{AdS}_5$ space, using gauge/string duality,
we have calculated the structure functions of the ``neutron''
which is dual to a spin-$\frac{1}{2}$ dilatino which is neutral corresponding to the $U(1)$ current we are considering.
We obtain both the unpolarized and polarized
structure functions. We find the structure functions of the ``neutron'' purely from Pauli interactions are power suppressed at
the same order as the ones of the ``proton'' purely from minimal interactions. We also
find that the Burkhardt-Cottingham sum rule for $g_2$ which is satisfied independent of $\tau$ and $q^2$
in the minimal interaction is broken due to  such a Pauli interaction term.

\begin{acknowledgments}
We acknowledge   fruitful discussions with Bo-Wen Xiao.  J.~Gao
acknowledges financial support by the China Postdoctoral Science Foundation
funded project under Contract No. 20090460736.
Z.~Mou is supported  by the National Natural
Science Foundation of China under the project Nos. 10525523 and 10975092, the Department of Science and Technology
of Shandong Province.

\end{acknowledgments}

\begin{appendix}

\section{Pauli  term from Kaluza-Klein reduction}
 In this appendix, we will illustrate how a Pauli interaction term can be produced from Kaluza-Klein reduction of higher dimensional
 fermion-graviton  coupling.  Since it is only  an illustration, for simplicity, let us just consider the reduction from 6D to 5D.
We start with the simple example  of a left chiral fermion field in 6D with the sixth dimension $\xi$ compactified.
 The action reads
\begin{eqnarray}
\label{S}
S=\int d^{5}x d\xi \sqrt{-G}\ \left\{\alpha{\overline{\Psi}}_L E^{M\cal{A}}\Gamma _{\cal{A}}\left( \partial _{M}+\frac{1}{%
2}\Omega _{M}^{\cal{BC}}\Sigma _{\cal{BC}}\right) \Psi_L +\mathrm{H.C.}\right\}
\end{eqnarray}%
where we have suppressed the pure graviton self-interaction and
\begin{eqnarray}
& &\Psi_L=\frac{1}{2}(1+\Gamma_{7})\Psi,
\ \ G_{MN}=\eta_{\cal{A}\cal{B}}E^{\cal{A}}_{M}E^{\cal{B}}_{N},
\  \Sigma_{\cal{A}\cal{B}}=\frac{1}{4}(\Gamma_{\cal{A}}\Gamma_{\cal{B}}-\Gamma_{\cal{B}}\Gamma_{\cal{A}}).
\end{eqnarray}
It should be noted that the coefficient $\alpha$ can be generally complex.
In the following, we use the capital $G, E, \Omega$ in the six-dimension
space and the lower $g, e, \omega$  in the five-dimensional space.  We will use
indices $M,N, ...$ to denote all six general curved spacetime dimensions,
and $\cal{A},\cal{B},...$  refer to all six local inertial spacetime dimensions,
while $m,n,...$ denote five dimensions in $\mathrm{AdS}_5$ and $a,b,...$ refer to five
dimensions in the local flat five-dimensional   spacetime.

Following  the formalism of  Kaluza-Klein, we  write the vielbein field as \cite{Lykken:1996xt}
\begin{eqnarray}
\label{vielbein}
E^{\cal{A}}_{N}=\left(\begin{array}{cc}e^{a}_{n}&A_{n}\\0&1\end{array}\right),
\ \ \ \ \ E^{N}_{\cal{A}}=\left(\begin{array}{cc}e^{n}_{a}&-A_{a}\\0&1\end{array}\right),
\ \ \ \ \ E^{N \cal{A}}=\left(\begin{array}{cc}e^{na}&0\\-A^{a}&1\end{array}\right),
\end{eqnarray}
where $A_{b}=e^{n}_{b}A_{n}, A^a=\eta^{ab}A_b$ and we have  set the scalar field simply as 1 and neglected
the dimension of the field. All the components in Eq.(\ref{vielbein}) do not depend on the sixth dimension coordinate $\xi$.

We make the split of $\Gamma$-matrices \cite{Duff:1999rk} as
\begin{eqnarray}
\Gamma_{\cal{A}}=\left( \gamma_{a}\otimes\sigma_3,1\otimes\sigma_{i}\right)\,, \ \ a=0,1,2,3,5 \ \ \mathrm{and}\ \  i=1,2
\end{eqnarray}
where $\gamma_{a}$ and $\sigma_{i} (i=1,2,3)$ are the D=5 $\gamma$-matrices and usual Pauli matrices, respectively.
We also decompose the spinor field as
\begin{eqnarray}
\Psi(x,\xi)=\psi(x)\otimes\chi(\xi)
\end{eqnarray}
where  $\psi(x)$ is the spinor in D=5, and $\chi(\xi)$ is the spinor in D=2.

Using the identity
\begin{eqnarray}
& &\Gamma_{\cal A} \Sigma_{\cal BC}=\Sigma_{AB}\Gamma_{\cal C}
+\frac{1}{2}\eta_{\cal AB}\Gamma_{\cal C}-\frac{1}{2}\eta_{\cal BC}\Gamma_{\cal A},\\
& &\Omega^{\cal{CB}}_{M}=E^{\cal C}_N \nabla_M E^{N\cal{B}}=-E^{N\cal{B}}\nabla_M E_N^{\cal C},
\end{eqnarray}
and the specific expressions  in Eq.(\ref{vielbein}),
we can have
\begin{eqnarray}
E^{M\cal{A}}\Omega^{\cal{CB}}_{M}\Gamma_{\cal{A}}\Sigma_{\cal{BC}}
&=&e^{ma}\omega^{cb}_{m}\Gamma_{a}\Sigma_{bc}
-e^{ma}e^{nb}\partial_m A_n\Sigma_{ab}\Gamma_{6}.
 \end{eqnarray}
Inserting it into the Lagrangian in Eq.(\ref{S}), we can obtain
\begin{eqnarray}
& &{\overline{\Psi}}_L E^{M\cal{A}}\Gamma _{\cal{A}}\left( \partial _{M}
+\frac{1}{2}\Omega _{M}^{\cal{BC}}\Sigma _{\cal{BC}}\right) \Psi_L\nonumber\\
&=&\left\{{\bar{\psi}} e^{ma}\gamma _{a}\left( \partial _{m}-i{\cal{Q}}A_m
+\frac{1}{2}\omega^{bc}_m\Sigma_{bc}\right)\psi
+i{\bar{\psi}}\left(i{\cal{Q}}+\frac{1}{4}e^{ma}e^{nb}F_{mn}\Sigma_{ab}\right)\psi
\right\}\nonumber\\
& &\times\left\{\chi^\dag\frac{(1+\sigma_2)}{2}\chi\right\}
\end{eqnarray}%
where we have taken the fermion to be in a charge eigenstate,
\begin{eqnarray}
\partial_6 \chi(\xi)=\frac{\partial}{\partial \xi}\chi(\xi)=i\cal{Q} \chi(\xi).
\end{eqnarray}
In such a way, the action can be rewritten  as,
\begin{eqnarray}
S&=&\int d^{5}x\sqrt{-g}\left\{\alpha\bar{\psi }e^{ma}\gamma _{a}\left( \partial _{m}-i{\cal Q}A_{m}
+\frac{1}{2}\omega _{m}^{bc}\Sigma_{bc}\right)\psi+\mathrm{H.C.}\right\}\nonumber\\
& &+ \int d^{5}x\sqrt{-g}
\left\{i\alpha\bar{\psi}\left( i{\cal Q}+\frac{1}{4}e^{mb}e^{nd}F_{mn}\Sigma_{bd}\right)\psi
+\mathrm{H.C.}\right\}
\end{eqnarray}%
where we have normalized
\begin{eqnarray}
\int d\xi \ \chi^\dag(y){(1+\sigma_2)}\chi(y)=1.
\end{eqnarray}%
Now we can see that  the Pauli interaction term in $\mathrm{AdS}_5$ space has been produced from higher 6D fermion-graviton interaction
by using Kaluza-Klein reduction.  For the uncharged fermion where ${\cal Q}=0$, only the Pauli interaction
term will contribute.  The reduction from 10D to 5D will be  similar except for more possible complications involved dealing
with more extra dimensions, which is beyond the scope of our present work.
\end{appendix}

\end{document}